%% file: main.tex
\documentclass[10pt]{sigplanconf}

\input{def/fonts}
\input{def/packages}

\input{def-import/macros}

\input{def/listings}
\input{def/macros}
\input{def-import/wavey.tex}

\makeatletter
\def\@copyrightspace{\relax}
\makeatother

\begin{document}

\title{Liveness for Verification\vspace{-1ex}}
\authorinfo{Roly Perera\and Simon J.~Gay}{School of Computing Science,
  University of Glasgow, UK}{\{roly.perera, simon.gay\}@glasgow.ac.uk\vspace{-6ex}}

\maketitle

\input{abstract}
\input{body}

{\footnotesize \bibliographystyle{abbrv}
\bibliography{main.bib}}

\end{document}

%% file: def/fonts.tex
\edef\keptrmdefault{\rmdefault}
\edef\keptsfdefault{\sfdefault}
\edef\keptttdefault{\ttdefault}
\usepackage[math]{iwona}
\edef\rmdefault{\keptrmdefault}
\edef\sfdefault{\keptsfdefault}
\edef\ttdefault{\keptttdefault}

\usepackage[T1]{fontenc}
\usepackage{libertine}
\usepackage[scaled=0.75]{beramono}
\usepackage{upgreek}

%% file: def/packages.tex
\usepackage{amsmath}
\usepackage{breakurl}
\usepackage{calc} 
\usepackage[usenames,dvipsnames]{color}
\usepackage{etoolbox}
\usepackage{float}
\usepackage[hidelinks]{hyperref}
\usepackage{listings}
\usepackage{microtype}
\usepackage{pifont} 
\usepackage[normalem]{ulem}
\usepackage{url}
\usepackage{xcolor}

\input{def-import/MnSymbol-triangles}

%% file: def-import/MnSymbol-triangles.tex
\DeclareFontFamily{U}{MnSymbolC}{}
\DeclareFontShape{U}{MnSymbolC}{m}{n}{
    <-6>  MnSymbolC5
   <6-7>  MnSymbolC6
   <7-8>  MnSymbolC7
   <8-9>  MnSymbolC8
   <9-10> MnSymbolC9
  <10-12> MnSymbolC10
  <12->   MnSymbolC12}{}
\DeclareFontShape{U}{MnSymbolC}{b}{n}{
    <-6>  MnSymbolC-Bold5
   <6-7>  MnSymbolC-Bold6
   <7-8>  MnSymbolC-Bold7
   <8-9>  MnSymbolC-Bold8
   <9-10> MnSymbolC-Bold9
  <10-12> MnSymbolC-Bold10
  <12->   MnSymbolC-Bold12}{}
\DeclareSymbolFont{MnSyC}{U}{MnSymbolC}{m}{n}

\DeclareMathSymbol{\smalltriangleright}{\mathord}{MnSyC}{72}
\DeclareMathSymbol{\smalltriangleup}{\mathord}{MnSyC}{73}
\DeclareMathSymbol{\smalltriangleleft}{\mathord}{MnSyC}{74}
\DeclareMathSymbol{\smalltriangledown}{\mathord}{MnSyC}{75}
\DeclareMathSymbol{\filledtriangleright}{\mathord}{MnSyC}{76}
\DeclareMathSymbol{\filledtriangleup}{\mathord}{MnSyC}{77}
\DeclareMathSymbol{\filledtriangleleft}{\mathord}{MnSyC}{78}
\DeclareMathSymbol{\filledtriangledown}{\mathord}{MnSyC}{79}

%% file: def-import/macros.tex
\newcommand*{\kw}[1]{{\text{\tt{#1}}}} 

\DeclareSymbolFont{bbsymbol}{U}{bbold}{m}{n}
\DeclareMathSymbol{\bbsemi}{\mathbin}{bbsymbol}{"3B}

\newcommand*{\figref}[1]{Figure~\ref{fig:#1}}

\newcommand*{\secref}[1]{Section~\ref{sec:#1}}

\newcommand*{\Secref}[1]{\S\,\ref{sec:#1}}

\newenvironment{nop}{}{}
\newenvironment{sdisplaymath}{
\begin{nop}\small\begin{displaymath}}{
\end{displaymath}\end{nop}\ignorespacesafterend}

\newenvironment{mathfig}{\begin{sdisplaymath}}{\end{sdisplaymath}}

\makeatletter
\newbox\sf@box
  {\def\sf@one{#1}%
   \def\sf@two{#2}%
   \setbox\sf@box\hbox
     \bgroup}%
  { \egroup
   \ifx\@empty\sf@two\@empty\relax
     \def\sf@two{\@empty}
   \fi
   \ifx\@empty\sf@one\@empty\relax
     \subfloat[\sf@two]{\box\sf@box}%
   \else
     \subfloat[\sf@one][\sf@two]{\box\sf@box}%
   \fi}
\makeatother

\definecolor{highlightcolor}{rgb}{1.0,0.8,0.8}
\definecolor{shadecolor}{rgb}{0.9,0.9,0.9}
\definecolor{lightgray}{rgb}{0.8,0.8,0.8}

\newenvironment{nscenter}
 {\parskip=0pt\par\nopagebreak\centering}
 {\par\noindent\ignorespacesafterend}

\newsavebox{\vardisplaymathbox}

%% file: def/listings.tex
\definecolor{verylightgray}{gray}{0.9}
\definecolor{lightgray}{gray}{0.5}
\definecolor{mediumgray}{gray}{0.45}
\newlength\lsthorizontalpadding
\newcommand*\lstnumberstyle{\ttfamily\scriptsize\textcolor{lightgray}}
\newlength\lstnumbersep
\newlength\lstnumberwidth
\lstdefinelanguage{code}{%
   morekeywords={system,behaviour,obj,using}%
  ,morecomment=[s]{\{-}{-\}}%
}
\lstnewenvironment{codecol}[1][]{\lstset{language=code,#1,moredelim=*[is][\textcolor{mediumgray}]{|}{|}}}{}

%% file: def/macros.tex
\newcommand*{\sendTriangle}{\filledtriangleleft}
\newcommand*{\receiveTriangle}{\filledtriangleright}

\makeatletter
\def\redwave{\bgroup \markoverwith{\lower3\p@\hbox{\sixly \textcolor{red}{\char58}}}\ULon}
\def\bluewave{\bgroup \markoverwith{\lower3\p@\hbox{\sixly \textcolor{blue}{\char58}}}\ULon}
\font\sixly=lasy6 
\makeatother
\newcommand\reduline{\bgroup\markoverwith{\textcolor{red}{\rule[-0.5ex]{2pt}{0.4pt}}}\ULon}
\newcommand\blueuline{\bgroup\markoverwith{\textcolor{blue}{\rule[-0.5ex]{2pt}{0.4pt}}}\ULon}

\newcommand*{\receivebad}[1]{\smash{\texttt{\blueuline{#1}}}}
\newcommand*{\sendbad}[1]{\smash{\texttt{\reduline{#1}}}}

\renewcommand*{\secref}[1]{\Secref{#1}} 

%% file: def-import/wavey.tex
\usepackage{xparse}
\usepackage{marginnote}
\usepackage{soul}
\usepackage{lipsum}
\usepackage{tikz}
\usetikzlibrary{calc}
\usetikzlibrary{decorations.pathmorphing}

\newlength\LineWidth
\newlength\Amplitude
\newlength\SegLength

\definecolor{HLcolor}{RGB}{240,0,0}

\newcommand\tikzmark[1]{%
  \tikz[overlay,remember picture] \node (#1) {};}

\makeatletter
\newcommand{\highlight@DoHighlight}{
  \draw[HLcolor,line width=\LineWidth,decorate,decoration={zigzag,amplitude=\Amplitude,segment length=\SegLength}]  ($(begin highlight)+(0,-2pt)$) -- ($(end highlight)+(0,-2pt)$) ;
}

\newcommand{\highlight@BeginHighlight}{
  \coordinate (begin highlight) at (0,0) ;
}

\newcommand{\highlight@EndHighlight}{
  \coordinate (end highlight) at (0,0) ;
}

\newdimen\highlight@previous
\newdimen\highlight@current

\DeclareRobustCommand*\highlight[1][]{%
  \SOUL@setup
  \def\SOUL@preamble{%
    \begin{tikzpicture}[overlay, remember picture]
      \highlight@BeginHighlight
      \highlight@EndHighlight
    \end{tikzpicture}%
  }%
  \def\SOUL@postamble{%
    \begin{tikzpicture}[overlay, remember picture]
      \highlight@EndHighlight
      \highlight@DoHighlight
    \end{tikzpicture}%
  }%
  \def\SOUL@everyhyphen{%
    \discretionary{%
      \SOUL@setkern\SOUL@hyphkern
      \SOUL@sethyphenchar
      \tikz[overlay, remember picture] \highlight@EndHighlight ;%
    }{%
    }{%
      \SOUL@setkern\SOUL@charkern
    }%
  }%
  \def\SOUL@everyexhyphen##1{%
    \SOUL@setkern\SOUL@hyphkern
    \hbox{##1}%
    \discretionary{%
      \tikz[overlay, remember picture] \highlight@EndHighlight ;%
    }{%
    }{%
      \SOUL@setkern\SOUL@charkern
    }%
  }%
  \def\SOUL@everysyllable{%
    \begin{tikzpicture}[overlay, remember picture]
      \path let \p0 = (begin highlight), \p1 = (0,0) in \pgfextra
        \global\highlight@previous=\y0
        \global\highlight@current =\y1
      \endpgfextra (0,0) ;
      \ifdim\highlight@current < \highlight@previous
        \highlight@DoHighlight
        \highlight@BeginHighlight
      \fi
    \end{tikzpicture}%
    \the\SOUL@syllable
    \tikz[overlay, remember picture] \highlight@EndHighlight ;%
  }%
  \SOUL@
}
\makeatother

\DeclareDocumentCommand\Wavey{O{red}O{1pt}O{5pt}m}{%
  \colorlet{HLcolor}{#1}
  \setlength\Amplitude{#2}%
  \setlength\SegLength{#3}%
  \tikzmark{endquote}\tikzmark{beginquote}\highlight{#4}%
}

%% file: abstract.tex
\begin{abstract}
  We explore the use of liveness for interactive program verification
  for a simple concurrent object language. Our experimental IDE
  integrates two (formally dual) kinds of continuous testing into the
  development environment: \emph{compatibility-checking}, which verifies
  an object's use of other objects, and \emph{compliance-checking},
  which verifies an object's claim to refine the behaviour of another
  object. Source code errors highlighted by the IDE are not static type
  errors but the reflection back to the source of runtime errors that
  occur in some execution of the system. We demonstrate our approach,
  and discuss opportunities and challenges.
\end{abstract}

%% file: body.tex
\section{Video submission}

This accompanies the video submission at
\url{https://vimeo.com/163716766}. We recommend reading the paper before
watching the video.

\section{Liveness for verification}

Liveness \cite{tanimoto90, hancock03, mcdirmid07} is often used to
provide a tight feedback loop between program output and edits, reducing
the cognitive burden on the programmer and supporting a more exploratory
development style. In the present work, we use liveness to provide a
tight feedback loop between \emph{runtime errors} and edits, as a form
of automated testing for concurrent programming that we call
\emph{language-integrated verification}. Our approach is related to
\emph{continuous testing} \cite{saff04,madeyski13}, which runs tests
automatically in the background and provides immediate feedback on test
failures.

We explore this idea in a language based on \emph{actors}
\cite{hewitt73}, concurrent objects that in response to a message can
explicitly transition to a new state offering different services. (From
now on by the term ``object'' we mean concurrent object in the actor
style.) The specific concurrency model is borrowed from communicating
automata~\cite{brand83}: objects communicate asynchronously, maintain a
separate FIFO mailbox for each client, and in every non-terminal state
are either sending to or receiving from a unique other object. Further
details on the language, programming model, and related work can be
found in \cite{perera16b}.

Language-integrated verification re-executes the program after an edit
to revalidate its behaviour, rather than recompute its output. We
explore the state space exhaustively, in the presence of
non-determinism, performing two model-checking style analyses.
\emph{Compatibility-checking} verifies that objects can be safely
composed, namely that every request for an interaction is eventually
honoured. \emph{Compliance-checking} verifies that when one system of
objects is declared to refine the observable behaviour of another, every
interaction supported by the refined system is supported by the refining
system.

These analyses are based on \emph{multiparty compatibility}, a notion
from communicating automata and session types
\cite{bocchi15,carbone15,denielou13}. However, whereas multiparty
compatibility is normally used to reason about \emph{types}, here we
apply it to objects. In our language there are no types, classes or
interfaces; instead any concrete object or system of objects can serve
as a specification of the behaviour of another. We describe our approach
in \secref{demo}, with reference to the accompanying demo, and discuss
limitations and future directions in \secref{conclusion}.

\input{fig/example/conf}

\section{Overview of demo}
\label{sec:demo}

As an example we model the interaction between a program committee and
an author during a conference submission. On the left of
\figref{example:conf} overleaf, we define a system called \kw{conf} with
a single object \kw{PC}; the blue underlining can be ignored for the
moment. A \emph{system} is simply one or more objects in parallel
composition. The \kw{author} object is left undefined; this is indicated
by the name appearing in italics. The PC expects the author to submit a
document, and then \emph{non-deterministically chooses} between sending
either \kw{reject} or \kw{conditionalAccept} back to the author. The
direction of the triangle means either blocking receive
($\receiveTriangle$) or non-blocking send ($\sendTriangle$);
non-singleton choices are enclosed in braces \kw{\{}\ldots\kw{\}}. A
period denotes a terminal state.

Explicit non-determinism is a non-standard language feature which serves
two important roles in our setting: it allows a single program or test
case to capture multiple scenarios, and it allows a concrete object to
be sufficiently abstract to serve as a specification. Whichever decision
is made by the PC, a string of review comments is returned to the
author. Here \kw{string} denotes not a \emph{type} but a
\emph{prototypical value} representing an unspecified string, consistent
with our typeless approach. (A refinement of this system might choose to
supply a concrete string instead.)

If the submission is accepted, the process enters an iterative phase:
the author submits further revisions until the paper is either
unconditionally accepted, or rejected. The iteration is implemented
using a \kw{\textbf{behaviour}} definition, which is simply a way of
giving a name to a state. The state \kw{Loop} is used twice here:
recursively in the body of \kw{revise}, and also immediately after the
definition of \kw{Loop}, as the body of \kw{conditionalAccept}. If the
paper is eventually accepted, an artifact request is issued, which the
author may decline, or respond to by providing a URL pointing to the
artifact.

We now discuss our two automated verification features:
compliance-checking (\secref{demo-overview:compliance}) and
compatibility-checking (\secref{demo-overview:compatibility}). These
involve executing all possible paths of the system and verifying that
every reachable configuration is good. Compliance-checking is formally
dual to compatibility-checking: to comply with an object (\emph{qua}
behavioural specification) is to be compatible with its dual, where one
dualises an object by turning sends into receives and vice versa.

\subsection{Compliance-checking}
\label{sec:demo-overview:compliance}

\figref{example:conf} illustrates compliance-checking, where we verify
that one system has the observable behaviour of another. On the right,
the programmer defines a new system \kw{conf'} which uses the colon
syntax shown to declare that it implements \kw{conf}. A number of
compliance errors are detected in various states and reflected back to
the relevant part of the source code. A convention we adopt for
visualising errors is that they are shown from the vantage point of the
system which has the focus, in this case \kw{conf'}.

Thus, the blue underlining on \kw{decline} that we disregarded earlier
reflects a state in \kw{conf} where the PC accepts a \kw{decline}
message from the author which the corresponding state in \kw{conf'} does
not support. The underlining of \kw{decline} in \kw{conf} should be
understood as a convenient way of indicating its \emph{absence} from
\kw{conf'}; other approaches are certainly possible. Dually, the red
underlining of \kw{accept} in \kw{conf'} reflects a state where the PC
sends an \kw{accept} message to the author that the corresponding state
in \kw{conf} does not permit.

Finally, the red underlining on the name \kw{artifact} reflects a state
requiring an interaction with the object \kw{artifact}, whereas the
corresponding state in \kw{conf} is terminal, as indicated by the period
following \kw{provide(URL)}.

\input{fig/example/author}

\subsection{Compatibility-checking}
\label{sec:demo-overview:compatibility}

For compatibility-checking, we verify that the objects in a system
compose in a safe way. In this example the programmer is able to build
the \kw{author} object interactively, using the compatibility errors to
guide the implementation. As part of the implementation, we introduce
another object, \kw{coauthor} (left undefined), which the author
consults in order to decide how to proceed if the paper is conditionally
accepted.

\figref{example:author} shows an interim implementation, with errors
which are again relativised to the system with focus, in this case
\kw{author}. The red wavey underlining on \kw{reject} on the left
reflects a state in which the author can only handle \kw{revise} or
\kw{accept}, but the PC wants to \kw{reject}. The blue underlining on
\kw{revise} and \kw{accept} in \kw{author} reflect the same runtime
error, and is in effect complementary to the red underlining on
\kw{reject} in \kw{PC}.

The red error on \kw{withdraw} on the right and the blue error on
\kw{submit} can be understood in the same mutually complementary way,
but with the polarity reversed: the author is trying to send a
\kw{withdraw} message to the PC in a state where the PC will only accept
\kw{submit}. At present the UI does not make the connection between
complementary errors apparent.

\section{Conclusions and challenges}
\label{sec:conclusion}

We described a prototypical IDE where errors reported to the user are
not \emph{type errors} but \emph{runtime errors} that occur in some
reachable configuration of the system. The programmer works in the
context of an active \emph{system}, which is simply a set of objects
composed in parallel; some represent application components being
developed or tested, and others serve as ``mock objects'' or test cases
representing exemplar scenarios. The programmer is responsible for
defining each system to be small enough for exhaustive checking yet
representative enough to give her confidence that the application
feature it validates is correct.

In return, our implementation performs exhaustive checking automatically
and provides a formal guarantee that execution paths validated in the
IDE will execute correctly under an asynchronous semantics based on
message queues ``in the wild''. Moreover any execution path which is
valid for the system remains valid if an object is replaced by a
compliant refinement of that object. Formalising the metatheory
corresponding to these guarantees is work-in-progress.

The current implementation is naive: there are significant limitations
relating to the language (\secref{future-work:language}), verification
methods (\secref{future-work:verification}), programming model
(\secref{future-work:UI}) and scalability
(\secref{future-work:efficiency}) which we intend to address in the
future.

\subsection{Language features}
\label{sec:future-work:language}

Our language lacks local and dynamically allocated objects, making it
only suitable for toy examples. The formalism of communicating automata
is extended with dynamic allocation in \cite{bollig13}; we plan to adapt
multiparty compatibility to this setting. Another language feature we
consider essential is inheritance, which requires a coinductive
definition for communicating automata. Our compliance-testing is
analogous to Java \kw{implements}, rather than Java \kw{extends}.

\subsection{Verification methods}
\label{sec:future-work:verification}

Like testing in general, but in contrast to a type system, our
verification method is complete rather than sound: it potentially
generates false positives rather than false negatives. One possible
route to increased coverage, whilst staying faithful to our concrete,
execution-oriented approach, is symbolic execution: this would allow
individual tests to cover multiple executions, and (soundly) reduce the
number of states that require explicit checking. Symbolic execution may
also be needed to verify programs with free variables, a situation which
arises often in our approach but which we have not properly considered
yet.

\subsection{User interface and programming model}
\label{sec:future-work:UI}

Our prototypical IDE is based on a conventional text editor, with
execution errors projected onto the source code. This presents a
familiar user interface but one unsuited to the actual task, which is
understanding and debugging problematic configurations (execution
states). In future work, we plan to integrate a debugger with the
editor, so that clicking on an error jumps to the corresponding
problematic configuration, allowing the programmer to see what went
wrong.

We would also like to combine our use of liveness for error reporting
with liveness for visualising output. One idea would be to introduce a
primitive object into our language representing a console or drawing
canvas, and then treat the sequence of messages sent to that object as
the program's output. Since we already explore every possible execution
path for verification purposes, computing all possible outputs would
incur no additional cost.

\subsection{Efficient implementation}
\label{sec:future-work:efficiency}

Modern software development workflows, such as test-driven development
\cite{beck02}, are ``incremental'', in that they emphasise verifying
\emph{changes} to the program, rather than the whole program. We plan to
exploit this by applying techniques from \emph{incremental computation}
\cite{acar05, hammer15} to our compatibility-testing and
compliance-testing algorithms. For certain kinds of edit, we should be
able to incrementally update the analysis rather than recompute it from
scratch. This is probably essential if our analyses are to scale to
non-toy examples whilst remaining responsive enough for interactive use.

%% file: fig/example/conf.tex
\begin{figure*}[ht]
\begin{nscenter}
\begin{minipage}[t]{0.45\textwidth}
\small
\begin{lstlisting}
system conf

obj PC
$\emph{author}\receiveTriangle$submit(doc)
$\emph{author}\sendTriangle${
   reject(string).
   conditionalAccept(string)
      behaviour Loop
         $\emph{author}\receiveTriangle$submit(doc)
         $\emph{author}\sendTriangle${
            reject(string).
            revise(string)
               Loop
            accept
               $\emph{author}\sendTriangle$artifactReq
               $\emph{author}\receiveTriangle${
                  $\receivebad{decline}$.
                  provide(URL).
               }
         }
      Loop
}\end{lstlisting}
\end{minipage}%
\quad\quad\quad%
\begin{minipage}[t]{0.45\textwidth}
\small
\begin{lstlisting}
system conf': conf

obj PC
$\emph{author}\receiveTriangle$submit(doc)
$\emph{author}\sendTriangle${
   $\sendbad{accept}$.
   reject(string).
   conditionalAccept(string)
      behaviour Loop
         $\emph{author}\receiveTriangle${
            submit(doc)
               $\emph{author}\sendTriangle${
                  reject(string).
                  revise(string)
                     Loop
                  accept
                     $\emph{author}\sendTriangle$artifactReq
                     $\emph{author}\receiveTriangle${
                        provide(URL)
                           $\sendbad{artifact}\sendTriangle${
                              certify.
                              noCertify.
                           }
                     }
               }
         }
      Loop
}\end{lstlisting}
\end{minipage}
\end{nscenter}
\caption{Live compliance-checking}
\label{fig:example:conf}
\end{figure*}

%% file: fig/example/author.tex
\begin{figure*}[ht]
\begin{nscenter}
\begin{minipage}[t]{0.45\textwidth}
\small
\begin{lstlisting}[escapechar=£]
system conf

obj PC
£$\emph{author}\receiveTriangle$£submit(doc)
£$\emph{author}\sendTriangle$£{
   reject(string).
   conditionalAccept(string)
      behaviour Loop
         £$\emph{author}\receiveTriangle$\hspace{-0.7em}\Wavey[blue]{submit}£(doc)
         £$\emph{author}\sendTriangle$£{
            £\hspace{-0.7em}\Wavey{reject}£(string).
            revise(string)
               Loop
            accept
               £$\emph{author}\sendTriangle$£artifactReq
               £$\emph{author}\receiveTriangle$£{
                  decline.
                  provide(URL).
               }
         }
      Loop
}\end{lstlisting}
\end{minipage}%
\quad\quad\quad%
\begin{minipage}[t]{0.45\textwidth}
\small
\begin{lstlisting}[escapechar=£]
system author
using conf

obj author
PC£$\sendTriangle$£submit("my paper")
PC£$\receiveTriangle$£{
   reject(str)
      £\emph{coauthor}$\sendTriangle$£rejected.
   conditionalAccept(str)
      behaviour Revise
         PC£$\sendTriangle$£{
            submit(string)
               PC£$\receiveTriangle$£{
                  £\hspace{-0.7em}\Wavey[blue]{revise}£(str)
                     Revise
                  £\hspace{-0.7em}\Wavey[blue]{accept}£
                     PC£$\receiveTriangle$£artifactReq
                     PC£$\sendTriangle$£provide("http://myurl.com").
             }
         }
      £\emph{coauthor}$\sendTriangle$£consult(str)
      £\emph{coauthor}$\receiveTriangle$£{
         continue
            Revise
         withdraw
            PC£$\sendTriangle$\hspace{-0.7em}\Wavey{withdraw}£.
     }
}\end{lstlisting}
\end{minipage}
\end{nscenter}
\caption{Live compatibility-checking}
\label{fig:example:author}
\end{figure*}